\documentclass[prd,aps,nofootinbib,onecolumn,showpacs,10pt]{revtex4}
\usepackage{graphicx,epsfig}

\usepackage{epsf}
\usepackage{graphicx}

\begin{document}

\title{  \bf On the Mass and Decay Constant of $K_2^*(1430)$ Tensor Meson}
\author{ T. M. Aliev$^{*1}$,
K. Azizi$^{\dag2}$ ,V. Bashiry$^{\ddag3}$ \\
$^1$Department of Physics, Middle East Technical University,
 06531 Ankara, Turkey\\
$^2$Physics Division,  Faculty of Arts and Sciences, Do\u gu\c s
University,
 Ac{\i}badem-Kad{\i}k\"oy, \\ 34722 Istanbul, Turkey\\
$^3 $ Engineering Faculty, Cyprus International University Via
Mersin 10, Turkey \\$^*$ e-mail:taliev@metu.ed.tr\\$^{\dag}$
{e-mail:kazizi@dogus.edu.tr}\\$^{\ddag}$ email:bashiry@ciu.edu.tr}

\begin{abstract}

The mass and decay constant of the ground state strange tensor meson $K_2^*(1430)$ with $I(J^P)=1/2(2^+)$ is calculated using the  QCD sum rules method. The results are consistent with the experimental data. It is found that   SU(3)  symmetry breaking effect constitutes about $20^0/_0$ of the decay constant.

\end{abstract}
\pacs{ 11.55.Hx,  14.40.Aq}

\maketitle

\section{Introduction}

Investigation of the properties of the tensor mesons is an area where experimental data is well ahead of the theoretical works. In this area there are a little theoretical works devoted to the analysis of the main characteristics of the tensor mesons (especially strange tensor mesons) and their decay modes comparing with scaler, pseudo-scalar, vector and axial vector mesons. Therefore, theoretical calculations on the physical parameters of these mesons and their comparison with experimental data could give essential information about their nature. The light tensor mesons spectroscopy can also be useful for understanding of low energy QCD. The QCD sum rules method as one of the most powerful and applicable tools to hadron physics could play an important role in this respect. Hadrons are formed in a scale of energy where the perturbation theory fails and to study the physics of mesons, we need to use some non-perturbative approaches. The QCD sum rules approach \cite{svz,colangelo,braun,balitsky},  which enjoys two peculiar properties namely its foundation based on QCD Lagrangian and free of model dependent parameters,  as one of the most well established  non-perturbative methods has been applied widely to
hadron physics.

Present work is devoted to the analysis of mass and decay constant of the strange tensor $K_2^*(1430)$ with quantum numbers $I(J^P)=1/2(2^+)$ by means of the QCD sum rules. The $K_2^*(1430)$ tensor meson together with  the unflavored  $a_2(1320)$, $f_2(1270)$ and
$f_{2}'(1525)$ are building the groundstate $1^3P_2~q\bar q$ nonet, which are experimentally well established in contrast to the scalar mesons \cite{von}. Note that, the mass and decay constant of light unflavored tensor mesons have been calculated in \cite{aliev}. Our aim here is to estimate the order of the SU(3) flavor symmetry breaking effects.  Recently, the magnetic moments of these mesons have also been predicted in lattice QCD \cite{lee}. The layout of the paper is as follows: in the next section, the sum rules for the mass and decay constant of the ground state strange tensor meson is calculated. Section III encompasses our numerical predictions for the mass and leptonic decay constant of the $K_2^*(1430)$ tensor meson.

\section{Theoretical Framework}
In this section, we calculate the mass and decay constant of the strange tensor meson in the framework of the QCD sum rules. The following correlation function,   the main object in this approach,   is the starting point:
\begin{eqnarray}\label{correl.func.1}
\Pi _{\mu\nu,\alpha\beta}=i\int
d^{4}xe^{iq(x-y)}{\langle}0\mid T[j _{\mu\nu}(x)
\bar j_{\alpha\beta}(y)]\mid  0{\rangle},
\end{eqnarray}
where, $j_{\mu\nu}$ is the interpolating current of the tensor meson. This current in the following form is responsible for creating the ground state strange tensor $K_2^*(1430)$ meson with quantum numbers $I(J^P)=1/2(2^+)$ from the vacuum:
\begin{eqnarray}\label{tensorcurrent}
j _{\mu\nu}(x)=\frac{i}{2}\left[\bar s(x) \gamma_{\mu} \overleftarrow{\overrightarrow{D}}_{\nu}(x) d(x)+\bar s(x) \gamma_{\nu} \overleftarrow{\overrightarrow{D}}_{\mu}(x) d(x)\right].
\end{eqnarray}
The $\overleftarrow{\overrightarrow{D}}_{\mu}(x)$ denotes the derivative with respect to x acting on left and right. it is given as:
\begin{eqnarray}\label{derivative}
\overleftarrow{\overrightarrow{D}}_{\mu}(x)=\frac{1}{2}\left[\overrightarrow{D}_{\mu}(x)-\overleftarrow{D}_{\mu}(x)\right],
\end{eqnarray}
where,
\begin{eqnarray}\label{derivative2}
\overrightarrow{D}_{\mu}(x)=\overrightarrow{\partial}_{\mu}(x)-i\frac{g}{2}\lambda^aA^a_\mu(x),\nonumber\\
\overleftarrow{D}_{\mu}(x)=\overleftarrow{\partial}_{\mu}(x)+i\frac{g}{2}\lambda^aA^a_\mu(x),
\end{eqnarray}
and the $\lambda^a$ are the Gell-Mann matrices and $A^a_\mu(x)$ are the external (vacuum) gluon fields.
In Fock-Schwinger gauge, $x^\mu A^a_\mu(x)=0$, this field can be expressed directly in terms of the gluon field strength tensor as:
\begin{eqnarray}\label{gluonfield}
A^{a}_{\mu}(x)=\int_{0}^{1}d\alpha \alpha x_{\beta} G_{\beta\mu}^{a}(\alpha x)=\frac{1}{2}x_{\beta} G_{\beta\mu}^{a}(0)+\frac{1}{3}x_\eta x_\beta D_\eta G_{\beta\mu}^{a}(0)+...
\end{eqnarray}

After this remark let calculate the correlation function presented in Eq. (\ref{correl.func.1}). In QCD sum rules approach, this correlation function  is calculated in two ways:
\begin{itemize}
 \item  Phenomenological or physical part which is obtained by saturating the correlation function with a tower of mesons with the same quantum numbers as the interpolating current,
\item  QCD or theoretical side which is obtained considering the internal structure of the hadrons, namely quarks and gluons and their interactions with each other and also the QCD vacuum . In this side the correlation function is calculated in deep Euclidean region, $q^2\ll0$, via operator product expansion (OPE) where the short and long distance effects are separated. The former is calculated using the perturbation theory, whereas the latter is represented in terms of  vacuum expectation values of the operators having different mass dimensions.
\end{itemize}
The sum rules for decay constant of the ground state meson is obtained isolating it in phenomenological part , equating both representations of the correlation function and applying Borel transformation to suppress the contribution of the higher states and continuum.

First, let us start to compute the physical side. Inserting a complete set of $K_2^*(1430)$ state between the currents in Eq. (\ref{correl.func.1}),  setting $y=0$  and performing intergal over x we obtain:
\begin{eqnarray}\label{phen1}
\Pi _{\mu\nu,\alpha\beta}=\frac{{\langle}0\mid j _{\mu\nu}(0) \mid
K_2^*\rangle \langle K_2^*\mid J_{\alpha\beta}(0)\mid
 0\rangle}{m_{K_2^*}^2-q^2}
&+& \cdots,
\end{eqnarray}
 where $\cdots$ represents the higher states and continuum contributions.
 The matrix elements creating the hadronic states out of vacuum
  can be
 written in terms of the leptonic decay constant and polarization tensor as follows:
\begin{eqnarray}\label{lep}
\langle 0 \mid J_{\mu\nu}(0)\mid K_2^*\rangle=f_{K_2^*} m_{K_2^*}^3\varepsilon_{\mu\nu}.
\end{eqnarray}
Combining two above equations and performing summation over  polarization tensor  using the relation
\begin{eqnarray}\label{polarizationt1}
\varepsilon_{\mu\nu}\varepsilon_{\alpha\beta}=\frac{1}{2}T_{\mu\alpha}T_{\nu\beta}+\frac{1}{2}T_{\mu\beta}T_{\nu\alpha}
-\frac{1}{3}T_{\mu\nu}T_{\alpha\beta},
\end{eqnarray}
where,
\begin{eqnarray}\label{polarizationt2}
T_{\mu\nu}=-g_{\mu\nu}+\frac{q_\mu q_\nu}{m_{K_2^*}^2},
\end{eqnarray}
we obtain the following final representation of the correlation function in physical side:
\begin{eqnarray}\label{phen2}
\Pi _{\mu\nu,\alpha\beta}=\frac{f^2_{K_2^*}m_{K_2^*}^6}{m_{K_2^*}^2-q^2}\left\{\frac{1}{2}(g_{\mu\alpha}g_{\nu\beta}+g_{\mu\beta}g_{\nu\alpha})\right\}+\mbox{other structures}+...
\end{eqnarray}
where, we have kept only structure which contains a contribution of the tensor meson.

On QCD side, the correlation function in  Eq. (\ref{correl.func.1})  is calculated using the explicit expression of the interpolating current presented in  Eq. (\ref{tensorcurrent}). After contracting out all quark pairs using the Wick's theorem we obtain:
\begin{eqnarray}\label{correl.func.2}
\Pi _{\mu\nu,\alpha\beta}=\frac{-i}{4}\int
d^{4}xe^{iq(x-y)}\left\{Tr\left[S_s(y-x)\gamma_\mu\overleftarrow{\overrightarrow{D}}_{\nu}(x)
\overleftarrow{\overrightarrow{D}}_{\beta}(y)S_d(x-y)\gamma_\alpha\right]+\left[\beta\leftrightarrow\alpha\right]+\left[\nu\leftrightarrow\mu\right]+\left[\beta\leftrightarrow\alpha,\nu\leftrightarrow\mu\right]\right\}.
\end{eqnarray}
From this equation, it follows that for obtaining the correlation function from QCD side the expression of the light quark propagator is needed. The light quark propagator is obtained in  \cite{R9714,R9715}:
\begin{eqnarray}\label{Slight}
S_q(x-y) &=& S^{free} (x-y) - {\langle qq\rangle \over 12} \left[1 -i {m_q \over
4} (\not\!{x} -\not\!{y})\right] - {(x-y)^2 \over 192} m_0^2 \langle qq\rangle
\left[1 -i {m_q \over 6} (\not\!{x}- \not\!{y})\right] \nonumber \\
&-& i g_s \int_0^1 du \left\{
{(\not\!x-\not\!y) \over 16 \pi^2 (x-y)^2} G_{\mu\nu}[u(x-y)] \sigma^{\mu\nu}
- u (x-y)^\mu G_{\mu\nu}[u(x-y)] \gamma^\nu {i \over 4 \pi^2 (x-y)^2} \right. \nonumber \\
&-& \left. i {m_q \over 32 \pi^2} G_{\mu\nu}[u(x-y)] \sigma^{\mu\nu} \left[ \ln\left( -{(x-y)^2
\Lambda^2 \over 4} \right) + 2 \gamma_E \right] \right\}~,
\end{eqnarray}
where, $\Lambda$ is the scale parameter,  we choose
it at the factorization scale $\Lambda=0.5~GeV-1.0~GeV$  \cite{R9716}, and
\begin{eqnarray}\label{Sfree}
S^{free} (x-y) = {i (\not\!x- \not\!y)\over 2 \pi^2 (x-y)^4} - {m_q \over 4 \pi^2 (x-y)^2}~.
\end{eqnarray}
Now, we put the expression of the propagators and apply the derivatives with respect to x and y in  Eq. (\ref{correl.func.2}) and eventually set $y=0$. As a result, we obtain the following expression in coordinate space:
\begin{eqnarray}\label{correl.func.3}
\Pi _{\mu\nu,\alpha\beta}=\frac{-i}{16}\int
d^{4}xe^{iqx}\left\{Tr\left[\Gamma_{\mu\nu,\alpha\beta}\right]+\left[\beta\leftrightarrow\alpha\right]+\left[\nu\leftrightarrow\mu\right]+\left[\beta\leftrightarrow\alpha,\nu\leftrightarrow\mu\right]\right\}.
\end{eqnarray}
where,
\begin{eqnarray}\label{fonk}
\Gamma_{\mu\nu,\alpha\beta}&=&\left\{{-i \not\!x \over 2 \pi^2 x^4} - {m_s \over 4 \pi^2 x^2} - {\langle \bar ss\rangle \over 12} \left(1 +i {m_s \over
4} \not\!{x} \right) - {x^2 \over 192} m_0^2 \langle \bar ss\rangle
\left(1 +i {m_s \over 6} \not\!{x} \right)\right\}\gamma_\mu\left\{\frac{2i}{\pi^2}\left[\frac{\gamma_\beta x_\nu}{x^6}+\frac{\gamma_\nu x_\beta+ \not\!{x}\delta_\beta^\nu}{x^6}-\frac{6\not\!{x}x_\beta x_\nu}{x^8}\right]\right.\nonumber\\&+&\left.\frac{m_0^2\langle\bar dd\rangle}{96}\delta_\beta^\nu\right\}\gamma_\alpha-\left\{\frac{i}{2\pi^2}\left[\frac{\gamma_\beta }{x^4}-\frac{4\not\!{x}x_\beta }{x^6}\right]-\frac{m_sx_\beta}{2 \pi^2 x^4}+\frac{im_s\langle\bar ss\rangle}{48}\gamma_\beta+\frac{m_0^2 \langle \bar ss\rangle }{96}x_\beta(1 +i \frac{m_s\not\!x}{6})+\frac{im_0^2 \langle \bar ss\rangle m_sx^2}{1152}\gamma_\beta\right\}\gamma_\mu\nonumber\\&&\left\{\frac{i}{2\pi^2}\left[\frac{\gamma_\nu }{x^4}-\frac{4\not\!{x}x_\nu}{x^6}\right]-\frac{m_0^2 \langle \bar dd\rangle x_\nu}{96}\right\}\gamma_\alpha-\left\{\frac{i}{2\pi^2}\left[\frac{-\gamma_\nu }{x^4}+\frac{4\not\!{x}x_\nu }{x^6}\right]+\frac{m_sx_\nu}{2 \pi^2 x^4}-\frac{im_s\langle\bar ss\rangle}{48}\gamma_\nu-
\frac{m_0^2 \langle \bar ss\rangle }{96}x_\nu(1 +i \frac{m_s\not\!x}{6})\right.\nonumber\\&-&\left.\frac{im_0^2 \langle \bar ss\rangle m_sx^2}{1152}\gamma_\nu\right\}\gamma_\mu\left\{\frac{i}{2\pi^2}\left[\frac{-\gamma_\beta }{x^4}+\frac{4\not\!{x}x_\beta}{x^6}\right]+\frac{m_0^2 \langle \bar dd\rangle x_\beta}{96}\right\}\gamma_\alpha+\left\{\frac{2i}{\pi^2}\left[\frac{-\gamma_\nu x_\beta}{x^6}-\frac{\gamma_\beta x_\nu+ \not\!{x}\delta^\beta_\nu}{x^6}+\frac{6\not\!{x}x_\beta x_\nu}{x^8}\right]\right.\nonumber\\&+&\left.\frac{m_0^2\langle\bar ss\rangle}{96}\delta^\beta_\nu(1 +i \frac{m_s\not\!x}{6})-\frac{m_s}{2\pi^2}\left[\frac{\delta^\beta_\nu}{x^4}-\frac{4x_\beta x_\nu}{x^6}\right]+\frac{im_0^2 \langle \bar ss\rangle m_s}{1152}\left[x_\beta\gamma_\nu+x_\nu\gamma_\beta\right]\right\}\gamma_\mu\left\{{i \not\!x \over 2 \pi^2 x^4}  - {\langle \bar dd\rangle \over 12}  - {x^2 \over 192} m_0^2 \langle \bar dd\rangle
\right\}\gamma_\alpha.\nonumber\\
\end{eqnarray}
In calculations, we neglected the  d quark mass. The calculations also  show that the terms proportional to the
gluon field strength tensor and four quark operators are very small (see also \cite{aliev}) and therefore, we do not present those terms in our final expression. The next step is to perform the integral over x using:
\begin{eqnarray}\label{integ}
\int d^4x \frac{e^{iqx}}{(x^2)^n}=(-1)^n (-i)\frac{\pi^2}{\Gamma(n)}\int_0^\infty d\alpha \alpha^{n-3}e^{\frac{-\widetilde{q}^2}{4\alpha}},
\end{eqnarray}
where, $\sim$ denotes the Euclidean space. Now, we separate the coefficient of the structure $\frac{1}{2}(g_{\mu\alpha}g_{\nu\beta}+g_{\mu\beta}g_{\nu\alpha})$ form both sides of the correlation function and apply the  Borel transformation as:
\begin{eqnarray}\label{borel}
&&\hat{{\textbf{\textit{B}}}}e^{\frac{-\widetilde{q}^2}{4\alpha}}=\delta(\frac{1}{M^2}-\frac{1}{4\alpha}),\nonumber\\
&&\hat{{\textbf{\textit{B}}}}\frac{1}{m^2-q^2}=e^{-m^2/M^2},
\end{eqnarray}
where, $M^2$ is the Borel mass squared. Finally, we obtain the following sum rule for the leptonic decay constant of the strange tensor meson:
\begin{eqnarray}\label{sumrul}
f_{K_2^*}^2e^{-m_{K_2^*}^2/M^2}=\frac{1}{m_{K_2^*}^6}\left\{\frac{N_c}{160\pi^2}\int_0^{s_0}dss^2e^{-s/M^2}-\frac{m_s}{24}m_0^2 \langle \bar dd\rangle\right\},
\end{eqnarray}
where, $N_c=3$ is the number of color and $s_0$ is the continuum threshold. The mass of the strange tensor meson is obtained by taking derivative with respect to $-\frac{1}{M^2}$ from the both sides of the above sum rule and dividing by itself, i.e.,
\begin{eqnarray}\label{sumrul2}
m_{K_2^*}^2=\frac{\frac{N_c}{160\pi^2}\int_0^{s_0}dss^3e^{-s/M^2}}{\frac{N_c}{160\pi^2}\int_0^{s_0}dss^2e^{-s/M^2}-\frac{m_s}{24}m_0^2 \langle \bar dd\rangle},
\end{eqnarray}
\section{Numerical analysis}

Present section is devoted to the numerical analysis of the sum rules for the mass and decay constant of the ground state strange tensor meson. The main input parameters entering to the sum rules expressions are continuum threshold $s_0$, Borel mass parameter $M^2$, strange quark mass and quark condensates. In further analysis, we put $ \langle \bar dd(1~GeV) \rangle = -(1.65\pm0.15)\times10^{-2}~GeV^3$ \cite{BL},
$\langle \bar ss(1~GeV) \rangle  = 0.8 \langle \bar uu(1~GeV)\rangle$,  $ m_{s}(2~GeV)=(111 \pm 6)~MeV$ at  $\Lambda_{QCD}=330~MeV$ \cite{Dominguez}, $m_0^2(1~GeV) = (0.8\pm0.2)~GeV^2$
\cite{R9718}. The continuum threshold $s_0$ and Borel mass parameter $M^2$ are auxiliary parameters, hence the physical quantities should be independent of them. Therefore, we look for working regions at which the physical quantities are weakly dependent on these parameters. The continuum
threshold $s_{0}$ is not completely arbitrary and it is  related to the energy
of the first exited state. The working region for the continuum threshold is obtained to be  the interval $s_0=(3-3.8)~GeV^2$.
 The working region for the Borel mass parameter
are determined by the requirement that not only  the higher state and
continuum contributions are suppressed but also the contribution of the
highest order operator must be small. In our analysis, the working
region for the Borel parameter is found to be $ 1~ GeV^2 \leq
M^2 \leq 3~ GeV^2 $. The dependence of the mass and decay constant of the ground state strange tensor meson are presented in Figs 1 and 2, respectively.
\begin{figure}[h!]
\begin{center}
\includegraphics[width=7cm]{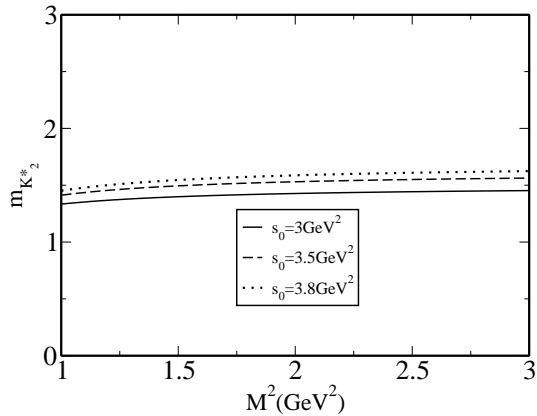}
\end{center}
\caption{The dependence of the mass of ground state strange tensor meson   on the Borel parameter $M^{2}$ at
three  fixed values of the continuum
threshold.} \label{fig1}
\end{figure}
\begin{figure}[h!]
\begin{center}
\includegraphics[width=7cm]{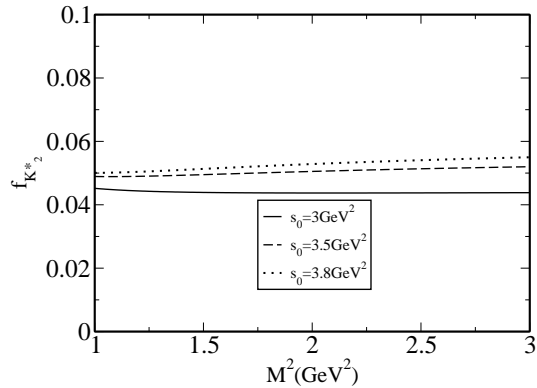}
\end{center}
\caption{The dependence of the decay constant of ground state strange tensor meson   on the Borel parameter $M^{2}$ at
three  fixed values of the continuum
threshold.} \label{fig2}
\end{figure}
Our final results on the mass and decay constant of the ground state strange meson are given as:
\begin{eqnarray}
m_{K_2^*}&=&(1.44\pm0.10)~GeV\nonumber\\
f_{K_2^*}&=&0.050\pm0.002
\end{eqnarray}
The result for the mass is in good consistency with the experimental value ($1.4321\pm0.0013$) \cite{pdg}. Comparing our results on the  decay constant of this meson  with predictions of the \cite{aliev} for the light unflavored tensor mesons,  we see that the SU(3)  Symmetry breaking effect is about $20^0/_0$.

\end{document}